\documentclass[aip,pop,floats,reprint]{revtex4-1}
\usepackage{amssymb}
\usepackage{amsmath}
\usepackage{graphicx}

 \def\bc{\begin{center}}          \def\ec{\end{center}}
 \allowdisplaybreaks
\begin{document}
 \title{Effect of beam emittance on self-modulation of long beams in plasma wakefield accelerators}
 \author{K.V.Lotov}
 \affiliation{Novosibirsk State University, 630090, Novosibirsk, Russia}
 \affiliation{Budker Institute of Nuclear Physics SB RAS, 630090, Novosibirsk, Russia}
 \date{\today}
 \begin{abstract}
The initial beam emittance determines the maximum wakefield amplitude that can be reached as a result of beam self-modulation in the plasma. The wakefield excited by the fully self-modulated beam decreases linearly with the increase of the beam emittance. There is a value of initial emittance beyond which the self-modulation does not develop even if the instability is initiated by a strong seed perturbation. The emittance scale at which the wakefield is twice suppressed with respect to the zero-emittance case (the so called critical emittance) is determined by inability of the excited wave to confine beam particles radially and is related to beam and plasma parameters by a simple formula. The effect of beam emittance can be observed in several discussed self-modulation experiments.
 \end{abstract}
 \maketitle

\section{Introduction}

The self-modulation instability\cite{EPAC98-806,PRL104-255003} (SMI) plays an important role in the concept of proton beam driven plasma wakefield acceleration.\cite{NatPhys9-363,PPCF56-084013} This instability transforms a long proton bunch into a train of short micro-bunches spaced one plasma wavelength apart and, therefore, enables efficient excitation of the plasma wave with state-of-the-art proton beams.\cite{PPCF53-014003,PoP18-103101} In turn, proton beams carry enough energy to drive the wave over hundreds of meters and accelerate electrons to energies of 1~TeV and beyond in a single stage.\cite{NatPhys9-363,PRST-AB13-041301,PoP18-103101,PRST-AB16-071301,SRep4-4171} This is the main advantage of proton drivers over electron and laser ones for which staging of many plasma sections is necessary to reach this energy scale.\cite{NIMA-410-388,NIMA-410-532,PRST-AB13-101301,PRST-AB14-091301,PRST-AB15-051301}

The SMI works properly only if a selected instability mode is somehow externally seeded.\cite{PRL104-255003,AIP1299-510,PRST-AB16-041301,PoP20-103111} Otherwise, non-axisymmetric perturbations (hosing modes) belonging to the same instability family\cite{PoP2-1326} could develop and destroy the beam. Even the axisymmetric modes are sufficient to destroy the beam, if several of them grow concurrently.\cite{NIMA-410-461,PPCF56-084014}

Another necessary condition of the long-distance beam propagation is a fine control of the instability with small variations of the plasma density along the beam path. In the perfectly uniform plasma, the seeded instability first transforms the long beam into the train of microbunches, and then heavily destroys them.\cite{PoP18-024501,PoP18-103101} However, if the beam encounters a small density increase (the plasma density step) at the linear stage of self-modulation, then the stable bunch train is formed which propagates in the plasma up to the beam energy depletion.\cite{PoP18-103101} Also, a gentle growth of the plasma density is favorable for trapping the witness electrons into optimal phases of the wave.\cite{Petrenko} Irregular or small scale density perturbations are, however, undesirable, as they are fatal for electrons accelerated by the excited wakefield.\cite{PoP19-010703,PoP20-013102}

By now, the properties of the SMI are known mostly from numerical simulations. The mechanism of bunch destruction in uniform plasmas and the role of the plasma density step are well understood, but for high quality beams only.\cite{smisim} Available analytical models are limited to narrow beams and the linear stage of the instability,\cite{PRL104-255003,PRL107-145003,PRL107-145002,AIP1507-103} with the primary attention given to the interplay of SMI and hosing modes.\cite{PRE86-026402,PoP20-056704,PRL112-205001} Experimental evidences of the SMI that come from electron or positron beams are too fragmentary for detailed comparison with the theory.\cite{PRL90-205002,PRL112-045001} The ultimate test of SMI properties will be possible with the AWAKE experiment\cite{PPCF56-084013,IPAC14-1537,NIMA-740-48,PoP21-123116} that is now under preparation at CERN. Several smaller scale experiments with electron beams\cite{NIMA-740-74,AIP1299-467,PoP19-063105,IPAC14-1476,AIP1507-559,AIP1507-594} are also discussed or prepared and may give valuable information.

In this paper, we numerically study the dependence of SMI properties on beam parameters, the most important of which is the beam emittance. Earlier studies of this kind\cite{PoP21-083107} were limited to the parameter area of interest for the AWAKE experiment at which the instability is mixed with many other effects. Here we study the SMI in the purest form and formulate universal limitations for it. In Sec.\ref{s2}, we define the problem under study and describe the test case that we take as an example. We then consider the uniform plasma in Sec.\ref{s3} and the plasma with the density step in Sec.\ref{s4}. In Sec.\ref{s5}, we discuss which of the proposals for experimental SMI studies may suffer from too large beam emittances.

\section{Statement of the problem} \label{s2}

We consider the same idealized setup as in Ref.\,[\onlinecite{smisim}] and use the same units of measure: speed of light $c$ for velocities, electron mass $m$ for masses, elementary charge $e$ for charges, initial plasma density $n_0$ for densities, inverse plasma frequency $\omega_p^{-1}$ for times, plasma skin depth $c/\omega_p$ for distances, and wavebreaking field $E_0 = mc\omega_p/e$ for fields. We also use cylindrical coordinates $(r, \varphi, z)$ and the co-moving coordinate $\xi = z - ct$.

Our aim is to induce general scalings from particular cases. For this, we simulate mono-energetic positron beams with the initial density profile
\begin{equation}\label{e1}
    n_b = \begin{cases}
        n_{b0} e^{-r^2/(2 \sigma_r^2)},\quad & -L < \xi<0, \\
        0, & \text{otherwise},
    \end{cases}
\end{equation}
which propagate in the positive $z$ direction through the cold, radially uniform plasma with immobile ions. Thus we exclude effects of ion motion\cite{PRL109-145005,PoP21-056705} and radial plasma non-uniformity\cite{PoP21-056703,PRL112-194801} from our consideration. We vary the relativistic factor $\gamma_b$, radius $\sigma_r$, density $n_{b0}$, length $L$, and emittance $\varepsilon_b$ of the beam in the neighborhood of some default values $\gamma_b = 1000$, $\sigma_r = 0.5$, $n_{b0} = 0.004$, $L=60$, and $\varepsilon_b = \varepsilon_{b0} = 4.65 \times 10^{-4}$. We intentionally consider only small beam densities to keep the plasma response linear and exclude effects of plasma non-linearity \cite{PoP20-083119,PoP21-083107} that may complicate the process.

We simulate the test cases with the axisymmetric quasi-static 2d3v hybrid code LCODE \cite{PoP5-785,LCODE}, in which the beam is modeled with macro-particles, and plasma is treated as the electron fluid.

\begin{figure}[tb]
\includegraphics[width=200bp]{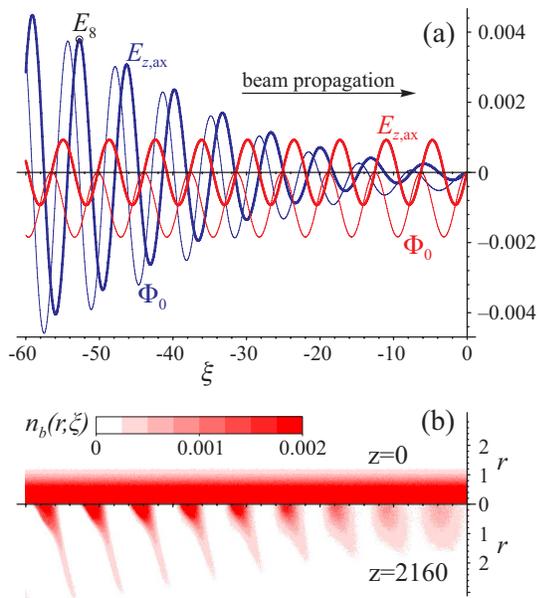}
\caption{(a) The on-axis electric field $E_{z,\text{ax}}$ (thick lines) and the depth of the radial potential well $\Phi_0$ (thin lines) at the beginning of interaction at $z=0$ (red) and at the moment of the strongest wakefield at $z=2160$ (blue) for the default beam parameters; (b) the beam density $n_b(r,\xi)$ at these propagation distances.}\label{f1-process}
\end{figure}
\begin{figure}[tbh]
\includegraphics[width=210bp]{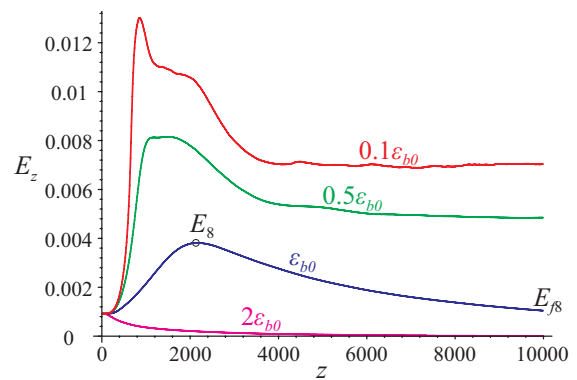}
\caption{The electric field at the 8th local maximum (inside the 8th micro-bunch) as a function of propagation distance for beams of various emittance. }\label{f2-growth}
\end{figure}

The process of beam self-modulation looks as follows. The hard leading edge leaves behind a seed wakefield [Fig.\,\ref{f1-process}(a)] that periodically focus or defocus beam slices. Perturbations of the beam radius and the plasma wave amplify each other and grow in space (in negative $\xi$ direction) and in time [Figs.\,\ref{f1-process}(a) and~\ref{f2-growth}]. Concurrently, the emittance driven divergence reduces the beam density at cross-sections of weak focusing [Fig.\,\ref{f1-process}(b)]. At the time of full beam bunching, the excited wakefield reaches its maximum value and then decreases due to partial destruction of the micro-bunches. The typical behavior of the wave amplitude at the 8th local maximum (inside the 8th micro-bunch) is shown in Fig.\,\ref{f2-growth} for several values of the initial beam emittance. Since the figure of merit for the self-modulation is the longitudinal field excited by the bunched beam, we focus our attention on the on-axis field $E_j$ at certain (numbered by the subscript $j$) local field maxima in $\xi$ and absolute maximum in $z$. This field maximum is located inside the $j$-th micro-bunch. The corresponding points for the default beam parameters are shown in Figs.\,\ref{f1-process}(a) and \ref{f2-growth} by small circles. As we see from Fig.\,\ref{f2-growth}, the wave amplitude does not always flatten out at large $z$. The amplitude $E_{fj}$ at long propagation distances $z=10000$ (which was analyzed in Ref.\,[\onlinecite{smisim}]) thus loses its significance for highly divergent beams, although it still contains some important information about the process.

\begin{figure}[b]
\includegraphics[width=187bp]{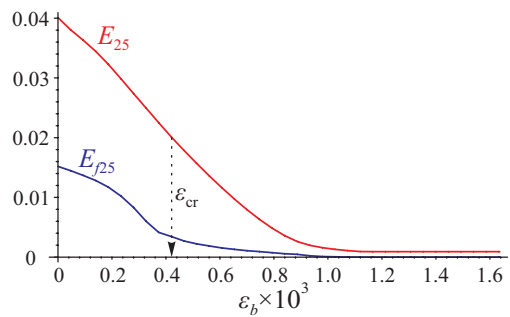}
\caption{The electric field at the distance of the strongest self-modulation ($E_{25}$) and at the long distance point ($E_{f25}$) as functions of the beam emittance for the default beam parameters. }\label{f3-typical}
\end{figure}

\section{Uniform plasma} \label{s3}

As we see from Fig.\,\ref{f2-growth}, the maximum wakefield excited by the self-modulating beam strongly depends on the initial beam emittance. This dependence is close to the linear one (Fig.\,\ref{f3-typical}) and can be conveniently characterized by the critical emittance $\varepsilon_\text{cr}$ that we define as the emittance at which the maximum field is reduced twice compared to the zero-emittance beam. The wave amplitude at long distances $E_{fi}$ drops down at the same emittance scale. The default emittance $\varepsilon_{b0}$ was deliberately chosen such that $\varepsilon_{b0} \approx \varepsilon_\text{cr}$ for $j=25$.

The ratio of the beam emittance to the critical emittance is thus an important beam feature that shows whether or not the particular beam can self-modulate efficiently. Let us relate the critical emittance to other (usually tabulated) beam parameters.

Two effects could be responsible for low fields at high emittances:\cite{PoP18-103101} beam dilution due to its divergence during development of the SMI and inability of the excited wave to confine beam particles radially. To analyze the first one, we estimate the time at which a beam particle shifts transversely by the distance of about $\sigma_r$; it is
\begin{equation}\label{e2}
    t_\text{diverg} \sim \sigma_r^2/\varepsilon_b.
\end{equation}
The characteristic time of SMI growth is\cite{PRL107-145003}
\begin{equation}\label{e3}
    t_\text{growth} \propto \sqrt{\frac{\gamma_b}{L n_{b0}}}.
\end{equation}
This expression differ from the commonly used ones\cite{PRL107-145003} in that we explicitly include the dependence on the beam length $L$ into the growth time. Equating the two times yields the scaling
\begin{equation}\label{e4}
    \varepsilon_\text{cr} \propto \sigma_r^2 \sqrt{\frac{L n_{b0}}{\gamma_b}}.
\end{equation}

For the second effect, we compare the energy of transverse motion of a beam particle\cite{PoP18-103101}
\begin{equation}\label{e5}
    W_\text{tr} = \frac{\varepsilon_b^2 \gamma_b}{2 \sigma_r^2}
\end{equation}
and the depth of the potential well $\Phi_0$ that we calculate more precisely than in Ref.\,[\onlinecite{PoP18-103101}]. If the beam density is factorable and has the form
\begin{equation}\label{e6}
    n_b (r, \xi) = n_{b0} f(r) g(\xi),
\end{equation}
then the force $\vec{F}$ acting on an axially moving relativistic positron is also factorable,
\begin{equation}\label{e7}
    \vec{F} \equiv \vec{E} + [\vec{e}_z, \vec{B}] = -\nabla \Phi (r, \xi) = -\nabla \bigl(n_{b0} F(r) G(\xi)\bigr),
\end{equation}
where $\vec{e}_z$ is the unit vector in $z$-direction, and expressions for $F(r)$ and $G(\xi)$ can be found, e.g., in Refs.\,[\onlinecite{PF30-252,IEEE-PS24-252,PoP12-063101}]. The density of the self-modulating beam is not factorable [Fig.\,\ref{f1-process}(b)], so the best estimate we can make for $\Phi_0$ is to assume that $G(\xi)$ oscillates with a linearly growing amplitude [as in Fig.\,\ref{f1-process}(a)], and $F(0)$ is the same as for the initial Gaussian beam:\cite{PoP12-063101}
\begin{equation}\label{e8}
    F(0) = \frac{\sigma_r^2}{2} e^{\sigma_r^2/2} \Gamma (0, \sigma_r^2/2),
\end{equation}
where
\begin{equation}\label{e9}
    \Gamma (\alpha, \beta) = \int_\beta^\infty t^{\alpha-1} e^{-t} dt.
\end{equation}
Collecting these expression together, we find
\begin{equation}\label{e10}
    \Phi_0 \propto \Phi_\text{sc} \equiv n_{b0} L \sigma_r^2 e^{\sigma_r^2/2} \Gamma (0, \sigma_r^2/2),
\end{equation}
and, by equating the amplitude of $\Phi_0$ and $W_\text{tr}$,
\begin{equation}\label{e11}
    \varepsilon_\text{cr} \propto \sigma_r^2 \sqrt{\frac{L n_{b0}}{\gamma_b} \, e^{\sigma_r^2/2} \Gamma (0, \sigma_r^2/2)}.
\end{equation}
If we consider the $j$-th field maximum, we take the coordinate $|\xi|$ of this maximum as $L$.

\begin{figure}[tb]
\includegraphics[width=194bp]{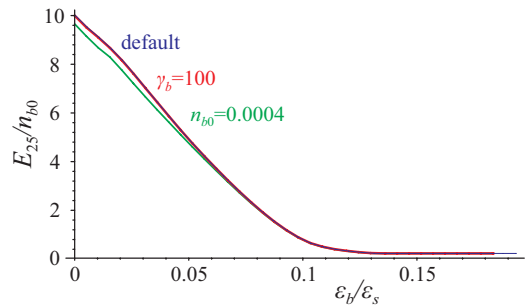}
\caption{The normalized dependence of the maximum accelerating field inside the $25$-th micro-bunch ($E_{25}$) on the beam emittance for the default parameter set (blue) and for beams of 10 times lower energy (red) or density (green).}\label{f4-scaling}
\end{figure}
\begin{figure}[b]
\includegraphics[width=214bp]{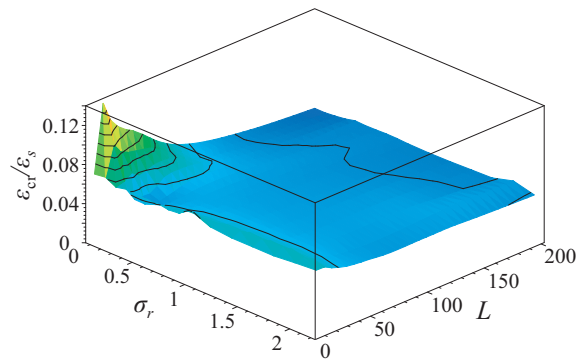}
\caption{Correctness of the scaling (\ref{e11}) for beams of various radius $\sigma_r$ and length $L$.}\label{f5-scaling}
\end{figure}
\begin{figure}[tbh]
\includegraphics[width=196bp]{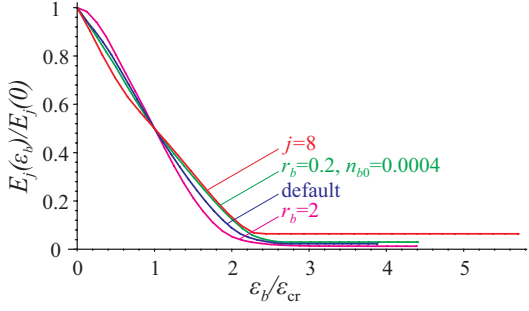}
\caption{The normalized dependence of the maximum accelerating field on the beam emittance for various beams. Beam parameters that differ from the default case ($25$-th local maximum) are indicated.}\label{f6-scaling}
\end{figure}

To check the correctness of the obtained scalings, we denote the right-hand side of (\ref{e11}) by $\varepsilon_s$ and compare it with simulation results. The scalings with beam energy $\gamma_b$ and density $n_{b0}$ are remarkably precise (Fig.\,\ref{f4-scaling}), and even the shape of the curve remains unchanged by order of magnitude variations of $\gamma_b$ and $n_{b0}$. The scalings with the beam radius $\sigma_r$ and length $L$ are correct everywhere except for short narrow beams (Fig.\,\ref{f5-scaling}). This figure also shows the numerical factor missed in derivation of (\ref{e11}): $\varepsilon_\text{cr} \approx 0.05\varepsilon_s$.

The scaling (\ref{e4}) is, obviously, incorrect. Consequently, the reduction of the maximum wakefield at high beam emittances happens because the wakefield of the self-modulated beam is not strong enough to confine all beam particles tightly focused. The wakefield reduction with respect to the idealized zero-emittance beam thus depends on the single parameter only, the ratio of the beam emittance to the critical emittance, in a wide interval of other beam parameters (Fig.\,\ref{f6-scaling}). If the beam particles are not positrons and have the mass $m_b$, then the critical emittance for this beam is
\begin{equation}\label{e12}
    \varepsilon_\text{cr} \approx 0.05 \sigma_r^2 \sqrt{\frac{L n_{b0}}{m_b \gamma_b} \, e^{\sigma_r^2/2} \Gamma (0, \sigma_r^2/2)}.
\end{equation}

\begin{figure}[tbh]
\includegraphics[width=180bp]{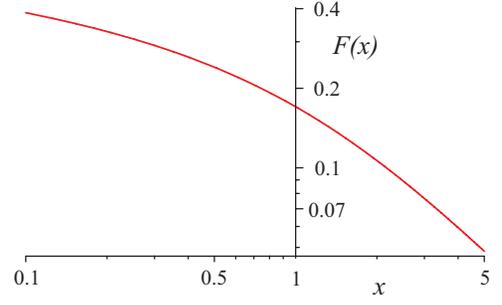}
\caption{The dimensionless function $F(x)$ for the critical beam divergence (\ref{e14}).}\label{f7-function}
\end{figure}

The formula (\ref{e12}) may be misleading in that it implicitly contains the plasma density through units of measure. In terms of the beam current $I_b = n_{b0} \sigma_r^2/2$, beam energy $W_b = m_b \gamma_b$, beam length measured in wave periods $j\approx L/(2 \pi)$, and in conventional units it is more informative:
\begin{equation}\label{e13}
    \varepsilon_\text{cr} \approx 0.18 \sigma_r \sqrt{j \frac{e I_b}{mc^3} \, \frac{mc^2}{W_b} \, e^{\omega_p^2 \sigma_r^2/(2 c^2)}\, \Gamma \!\! \left( 0, \frac{\omega_p^2 \sigma_r^2}{2 c^2}\right)}.
\end{equation}
Of importance is thus the initial angular spread of the beam $\delta \alpha = \varepsilon_b / \sigma_r$, the critical value for which is
\begin{equation}\label{e14}
   \alpha_\text{cr} \approx \sqrt{j \frac{e I_b}{mc^3} \, \frac{mc^2}{W_b}} \, F\!\left( \frac{\omega_p^2 \sigma_r^2}{2 c^2}\right)
\end{equation}
with the dimensionless function $F(x)$ shown in Fig.\,\ref{f7-function}. For the most interesting case of $\sigma_r \approx c/\omega_p$, $F(x) \approx 0.2$.
\begin{figure}[tbh]
\includegraphics[width=209bp]{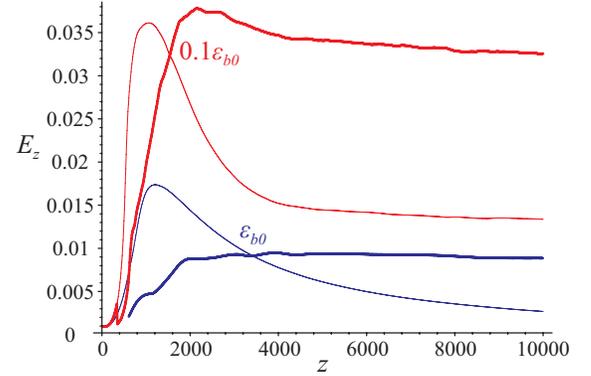}
\caption{The electric field at the 25th local maximum  (inside the 25th micro-bunch) as a function of propagation distance for beams of various emittance with (thick lines) and without (thin lines) the plasma density step. The step parameters are optimized for the maximum field at $z=10000$.}\label{f8-stepgr}
\end{figure}
\begin{figure}[tbh]
\includegraphics[width=198bp]{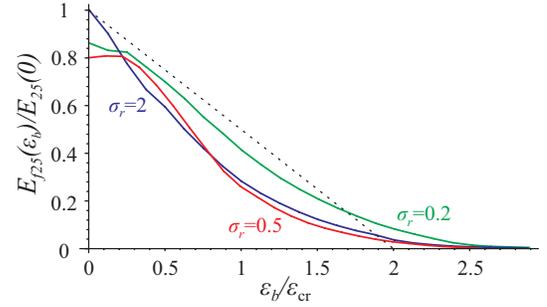}
\caption{The normalized dependence of the established accelerating field $E_{f25}$ on the beam emittance for the plasma with the optimized density step and beams of various radius (indicated on the plot). The field is normalized to the maximum field in the uniform plasma at zero emittance. The dash line shows the linear decrease typical for $E_j(\varepsilon_b)$.}\label{f9-steptyp}
\end{figure}
\begin{figure}[tbh]
\includegraphics[width=226bp]{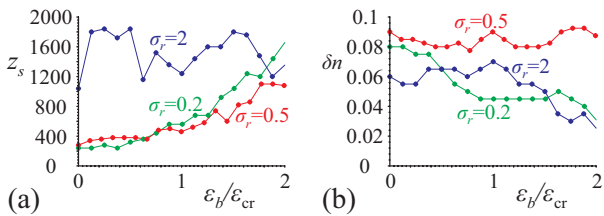}
\caption{Optimum location (a) and magnitude (b) of the density step as functions of the beam emittance for beams of different radius (indicated on the plots). The step parameters are optimized for the maximum $E_{f25}$. Points are obtained from simulations, connecting lines are added for visibility.}\label{f10-locstep}
\end{figure}
\begin{table*}[t]
 \caption{ Proposals for SMI experiments}\label{t1}
 \bc\begin{tabular}{llllllllll}\hline
  Facility, reference & $L$, cm \quad & $\sigma_r$, $\mu$m \quad & $I_b$, A \quad & $W_b$, GeV \quad & $\varepsilon_b$, mm mrad \quad & $n_0$, $\text{cm}^{-3}$ \quad & $j$ & $\varepsilon_b / \varepsilon_\text{cr}$ \quad & $A_\text{nl}$ \\ \hline
  PITZ (DESY, Zeuthen)\cite{NIMA-740-74} & 0.593 & 42 & 5 & 0.0215 & 0.009 & $10^{15}$ \quad & 5 & 0.07 & 0.05 \\
  FACET (SLAC, Stanford)\cite{PoP19-063105} & 0.1 & 10 & 770 & 20.5 & 0.0012 & $2.3\times 10^{17}$ & 14 & 0.12 & 1.6 \\
  LHC (CERN, Geneva)\cite{PoP18-103101} & 8 & 100 & 27 & 7000 & 0.0005 & $3\times 10^{15}$ & 130 & 0.17 & 0.4 \\
  AWAKE (CERN, Geneva)\cite{PoP21-123116} & 12 & 200 & 50 & 400 & 0.009 & $7\times 10^{14}$ & 95 & 0.3 & 0.5 \\
  IC VEPP-5 (BINP, Novosibirsk)\cite{AIP1299-467} & 1.6 & 20 & 60 & 0.511 & 0.023 & $5\times 10^{14}$ & 11 & 0.3 & 11 \\
  PS (CERN, Geneva)\cite{PPCF53-014003} & 40 & 400 & 10 & 24 & 0.08 & $5\times 10^{14}$ & 270 & 0.8 & 0.1 \\
  ATF (BNL, Brookhaven)\cite{AIP1507-559} & 0.096 & 120 & 310 & 0.0583 & 0.11 & $3.03\times 10^{16}$ \quad & 5 & 1.05 & 0.01 \\
  PROTOPLASMA (Fermilab, Batavia)\cite{AIP1507-644} \quad & 20 & 100 & 20 & 120 & 0.03 & $10^{16}$ & 600 \quad & 1.4 & 0.4 \\
  \hline
 \end{tabular}\ec
\end{table*}

\section{Plasma with a density step} \label{s4}

A small step-like increase $\delta n$ of the plasma density at the linear stage of SMI development results in formation of the bunch train that stably propagates in the plasma for a long distance\cite{PoP18-103101,smisim}. The established wave amplitude in this case also depends on the initial beam emittance (Fig.\,\ref{f8-stepgr}). The emittance scale for the twofold reduction of the established wakefield is the same, $\varepsilon_\text{cr}$, though the law of field reduction may differ from the linear one (Fig.\,\ref{f9-steptyp}). A large beam emittance also changes the optimum location $z_s$ of the step [Fig.\,\ref{f10-locstep}(a)]. The optimum location corresponds to the linear stage of the instability\cite{smisim}, so the slower wave growth observed at large radii or emittances of the beam results in later optimum steps. The optimum magnitude $\delta n$ does not show any significant trend [Fig.\,\ref{f10-locstep}(b)], as it is determined by the beam length $L$, which does not change.

\section{Available drive beams} \label{s5}

Let us apply the obtained formulae to beams that have been discussed in the context of self-modulation studies. The key parameters of proposed experiments are listed in Table~\ref{t1}. The parameter $L$ is the beam length for flattop beams, $L=2\sigma_z$ for Gaussian beams ($\sigma_z$ is the root-mean-square beam length), and $L=\sigma_z$ for proposals with half-cut beams (AWAKE, LHC). We see that several cases have $\varepsilon_b / \varepsilon_\text{cr} \gtrsim 1$, so the emittance driven erosion may suppress self-modulation there.

The last column of Table~\ref{t1} shows whether the above linear theory is applicable to these beams or not. As a quick measure of nonlinearity we take
\begin{equation}\label{e15}
    A_\text{nl} = j n_{b0}.
\end{equation}
This is the estimated perturbation of the plasma density by $j$ micro-bunches of the density $n_{b0}$ that resonantly drive the wave. The linear approximation violates at $A_\text{nl} \sim 0.5$ by the nonlinear elongation of the plasma wave\cite{PoP20-083119}. If $A_\text{nl} \gg 1$, then the beam-plasma interaction is strongly nonlinear, and hosing modes may dominate self-modulation even if the latter is properly seeded\cite{PRL112-205001}.

\acknowledgments

This work is supported by The Russian Science Foundation, grant No.~14-12-00043. The computer simulations are made at Siberian Supercomputer Center SB RAS.

\end{document}